# Landauer vs. Boltzmann and Full Dispersion vs. Debye Model Evaluation of Lattice Thermal Conductivity

Changwook Jeong, Supriyo Datta, and Mark Lundstrom

Network for Computational Nanotechnology
Birck Nanotechnology Center
Purdue University
West Lafayette, Indiana, 47907

**Abstract**- Using a full dispersion description of phonons, the thermal conductivities of bulk Si and $Bi_2Te_3$ are evaluated using a Landauer approach and related to the conventional approach based on the Boltzmann transport equation. A procedure to extract a well-defined average phonon mean-free-path from the measured thermal conductivity and given phonon-dispersion is presented. The extracted mean-free-path has strong physical significance and differs greatly from simple estimates. The use of simplified dispersion models for phonons is discussed, and it is shown that two different Debye temperatures must be used to treat the specific heat and thermal conductivity (analogous to the two different effective masses needed to describe the electron density and conductivity). A simple technique to extract these two Debye temperatures is presented and the limitations of the method are discussed.

## 1) Introduction

Electron and phonon transport play a critical role in a number of technological applications. They are central to thermoelectric technology, for which performance is determined by the dimensionless figure of merit, *ZT,* which has been limited to ~ 1 [1-3] for many years. Recent reports of *ZT >1*, [1-3] have been achieved by using nanostructured materials to suppress the lattice thermal conductivity. Further progress will require careful engineering of both the phonon and electron transport[4-9]. Phonon transport also plays an increasingly important role in integrated circuits where the increasing importance of power dissipation, self-heating, and the management of hot spots [10] necessitates electron-thermal co-design. These examples indicate that a unified treatment of electron and phonon transport would be useful.

Diffusive transport has been often described by the Boltzmann transport equation (BTE) and simplifications of it, such as drift-diffusion equation for electrons or Fourier's Law for phonons [11,12]. The Landauer approach[13] provides a simple, physically insightful description of ballistic transport and has been widely used to describe quantized electrical and thermal transport in nanostructures[14-17]. Although not as widely appreciated, the Landauer approach describes diffusive transport as well and provides a simple way to treat the ballistic to diffusive transition. Thermal transport in nanowires has recently been described by a Landauer approach[18], but applications of the Landauer approach to bulk transport have been rare. In a previous paper[19], we showed a very simple procedure to use the Landauer approach with a full band description of $E(k)$ to evaluate thermoelectric transport parameters. In that paper, we also related the full band calculations to the widely-used effective mass level model and presented a procedure for extracting the two different effective masses (density-of-states and conductivity effective masses) that are needed to evaluate the electron density and the TE transport coefficients.

In this paper, we extend Ref. 19 to phonons and show how the same very simple procedure can be used to evaluate the lattice thermal conductivity from a full zone description of the phonon dispersion. The main contributions of the paper consist of presenting a simple technique for extracting a physically meaningful mean-free-path for phonons from the measured thermal conductivity and relating the full dispersion results to the simpler, Debye model for phonon dispersion. Our specific objectives are: 1) to mathematically relate the Landauer expression for the thermal conductivity, $\kappa_{ph}$ to the more common approach that begins with classical kinetic theory, 2) to show that two different Debye temperatures are needed to accurately evaluate both the specific heat and lattice thermal conductivity with physically meaningful mean-free-paths, 3) to quantitatively examine both electronic performance and thermal performance of bulk silicon (Si) and bismuth telluride ($Bi_2Te_3$) within the Landauer framework and a full zone description of the phonon dispersion, 4) to present a technique to extract a clearly-defined average phonon mean-free-path for phonon backscattering from the measured thermal conductivity, and 5) to discuss the similarities and differences between electron and phonon transport in terms of the Landauer picture.

The paper is organized as follows. In Sec. 2, we present a brief summary of the Landauer formalism for electron and phonon transport. In Sec. 3, the results of full phonon dispersion simulations of the electrical and thermal conductivities of Si and $Bi_2Te_3$ are presented. A technique to extract a well-defined mean-free-path is presented in Sec. 4. We also compare and contrast electron and phonon transport in Sec. 4 and discuss the extraction of the two Debye temperatures that are needed when using simplified models of phonon dispersion to evaluate the specific heat and thermal conductivity. Finally, our conclusions are summarized in Sec. 5.

**2) Approach**

The theoretical approach to phonon transport used in this paper closely follows the approach for electrons as presented in Ref. [19]. In the linear response regime, we can define[20,21]

$$G(E) = \left(\frac{2q^2}{h}\right)\bar{T}_{el}(E) = \frac{A}{L}q^2\Sigma(E) \qquad [1/\Omega], \qquad (1)$$

where

$$\bar{T}_{el}(E) = T_{el}(E) M_{el}(E) = \frac{\lambda_{el}(E)}{\lambda_{el}(E) + L} M_{el}(E) \qquad (2)$$

is the total transmission for electrons in the Landauer picture[13] with $M_{el}(E)$ being the number of conduction channels at $E$ and $T_{el}(E)$ is the transmission at energy, $E$, with $\lambda_{el}(E)$ being the mean-free-path for backscattering and $L$ the length of the resistor. Equation (1) also expresses $G(E)$ in terms of the so-called transport distribution, $\Sigma(E)$, which arises when solving the diffusive Boltzmann transport equation (BTE)[21] and is defined as

$$\Sigma(E) = \frac{1}{\Omega} \sum_{\vec{k}} v_x^2(\vec{k}) \tau(\vec{k}) \delta(E - E(\vec{k})) \qquad (3)$$

where $\Omega = AL$ and $A$ is the cross-sectional area of the conductor.

Although the approach is more general, in this paper we restrict our attention to diffusive samples for which $T_{el}(E) \to \lambda_{el}(E)/L$ and to three dimensional samples for which we write $G(E) = \sigma(E) A/L$, where $\sigma(E)$ is the conductivity. Accordingly, the expressions for conductivity analogous to the conductance in Eq. (1) become

$$\sigma(E) = \left(\frac{2q^2}{h}\right) \frac{M_{el}(E)}{A} \lambda_{el}(E) = q^2 \Sigma(E) \qquad [1/\Omega\text{-m}], \qquad (4a)$$

The total electrical conductivity is obtained by integrating $\sigma(E)(-\partial f_0/\partial E)$ over all of the energy channels, and the other thermoelectric coefficients are readily obtained, as described in Ref. [19], where mathematical definitions of $M_{el}(E)$ and $\lambda_{el}(E)$ are also given. For example, the electronic thermal conductivity for zero voltage gradient is obtained as

$$\kappa_0 = \left(\frac{2k_B^2 T_L \pi^2}{3h}\right) \int_{-\infty}^{+\infty} \frac{M_{el}(E)}{A} \lambda_{el}(E) \left(\frac{3}{\pi^2}\left(\frac{E-E_F}{k_B T_L}\right)^2 \left(-\frac{\partial f_0}{\partial E}\right)\right) dE \qquad \text{[W/m-K]}. \qquad (4b)$$

Expressions for the lattice thermal conductance, $K_{ph}$, and lattice thermal conductivity, $\kappa_{ph}$ analogous to Eq. (4b) can be readily obtained and expressed as [14,22]

$$K_{ph}(\omega) = \left(\frac{k_B^2 T_L \pi^2}{3h}\right) \bar{T}_{ph}(\omega) \qquad \text{[W/K]} \qquad (5a)$$

and in the diffusive limit

$$\kappa_{ph}(\omega) = \left(\frac{k_B^2 T_L \pi^2}{3h}\right) \frac{M_{ph}(\omega)}{A} \lambda_{ph}(\omega) \qquad \text{[W/m-K]}, \qquad (5b)$$

where $n_0$ is the Bose-Einstein distribution function, the transmission is $\bar{T}_{ph}(\omega) = T_{ph}(\omega) M_{ph}(\omega)$ and $M_{ph}(\omega)$ is the number of phonon conducting modes (per polarization). The definitions of $T_{ph}(\omega)$ and $M_{ph}(\omega)$ are similar to those for electrons[19]. In 3D, the phonon mean-free-path for backscattering, $\lambda_{ph}(\omega)$ is given as [19]

$$\lambda_{ph}(\omega) = (4/3) \upsilon_{ph}(\omega) \tau_{ph}(\omega) = (4/3) \Lambda(\omega) \qquad (6)$$

where the pre-factor, 4/3 comes from averaging over angle in 3D, $\upsilon_{ph}(\omega)$ is the spectral phonon group velocity at frequency, $\omega$, $\tau_{ph}(\omega)$ is the phonon momentum relaxation time, and $\Lambda(\omega)$ is the commonly-defined spectral mean free path. Note that the mean-free-path for backscattering, Eq. (6), which arises in the Landauer approach, is somewhat longer than the mean-free-path for scattering, $\Lambda$. In the appendix, the relation of Eq. (5b), the Landauer expression for lattice thermal conductivity, to the conventional expression from kinetic theory is given.

To find the total conductivities, we multiply the energy-resolved quantities by a "window function" and integrate over energy,

$$\sigma = \left(\frac{2q^2}{h}\right)\int_{-\infty}^{+\infty}\frac{M_{el}(E)}{A}\lambda_{el}(E)\left(-\frac{\partial f_0}{\partial E}\right)dE \equiv \left(\frac{2q^2}{h}\right)\langle M_{el}\rangle\langle\langle\lambda_{el}\rangle\rangle \tag{7a}$$

$$\kappa_{ph} = \left(\frac{k_B^2 T_L \pi^2}{3h}\right)\int_{-\infty}^{+\infty}\frac{M_{ph}(\omega)}{A}\lambda_{ph}(\omega)\left(\frac{3}{\pi^2}\left(\frac{\hbar\omega}{k_B T_L}\right)^2\left(-\frac{\partial n_0}{\partial(\hbar\omega)}\right)\right)d(\hbar\omega)$$

$$\equiv \left(\frac{k_B^2 T_L \pi^2}{3h}\right)\langle M_{ph}\rangle\langle\langle\lambda_{ph}\rangle\rangle \tag{7b}$$

where

$$\langle M\rangle \equiv \int M(x)W(x)dx \tag{7c}$$

$$\langle\langle\lambda\rangle\rangle \equiv \langle M\lambda\rangle/\langle M\rangle = \int\lambda(x)M(x)W(x)dx\bigg/\int M(x)W(x)dx \tag{7d}$$

where $x = E$ for electrons and $x = \hbar\omega$ for phonons and $W(x)$ is a "window" function given by

$$W_{el}(E) = \left(-\frac{\partial f_0}{\partial E}\right) \quad\text{(electrons)} \tag{7e}$$

$$W_{ph}(\hbar\omega) = \frac{3}{\pi^2}\left(\frac{\hbar\omega}{k_B T_L}\right)^2\left(-\frac{\partial n_0}{\partial(\hbar\omega)}\right) \quad\text{(phonons)} \tag{7f}$$

The electrical conductivity is proportional to the quantum of electrical conductance, $2q^2/h$, and the thermal conductivity to the quantum of thermal conductance, $k_B^2 T_L \pi^2/3h$. The electrical and thermal conductivities are related to these two fundamental parameters and to the number of conducting channels per unit area, the mean-free-paths for backscattering, and to the Fermi-Dirac or Bose-Einstein distributions.

The number of conducting channels is determined by the electronic structure or phonon dispersion of the material. In Ref. [20], we discussed the evaluation of this quantity for electrons in 1D, 2D, and 3D considering a simple, effective mass level model. In Ref. [19], we discussed the evaluation of $M_{el}(E)$ from a full band description of $E(k)$ and its relation to effective mass level models. For phonons, a linear and isotropic phonon dispersion, $\omega = \upsilon_s q$, gives $M_{ph}(\omega)$ and $D_{ph}(\omega)$ as

$$M_{ph}(\omega) = A\left(3\omega^2/4\pi\upsilon_s^2\right) \qquad \text{(3D)} \qquad (8a)$$

$$D_{ph}(\omega) = \Omega\left(3\omega^2/\pi h\upsilon_s^3\right) \qquad \text{(3D)} \qquad (8b)$$

where $\upsilon_s$ is the velocity of sound in the direction of transport, and the factor of 3 comes from 3 branches. In the appendix, corresponding expressions for 1D, 2D, and 3D conductors are given and compared to the expressions for electrons. Our objective in this paper is to present a simple technique to compute, $M_{ph}(\omega)$ from a given full zone description of the phonon dispersion. Because simple descriptions of phonon dispersions are convenient to use, they find wide applications. The extraction of Debye model parameters from a rigorous evaluation of $M_{ph}(\omega)$ and a discussion of the limitations of the Debye model are, therefore, also important parts of this paper.

Given a phonon dispersion, $M_{ph}(\omega)$ can be obtained by counting the bands that cross the energy of interest. This method provides a computationally simple way to obtain $M_{ph}(\omega)$ from a given $\omega(q)$ [18,19]. The basic idea is illustrated in Fig. 2(a) using a dispersion relation along the transport direction for Si: 1) from 0 to 7.3 *THz*, the $M_{ph}(\omega)$ is three due to two transverse AP modes (TA) and one longitudinal AP modes (LA), 2) from 7.3 to 12.8 to 15.0 *THz*, we have only one LA and one longitudinal OP modes (LO). 3) from 15.0 to 15.8 *THz*, $M_{ph}(\omega)$ in this case is 3 due to two transverse OP modes (TO) and one LO. To evaluate $M_{ph}(\omega)$, a full band description of phonon dispersion is needed. Several techniques have been reported for computing detailed phonon bandstructure[23,24]. In this work, the full phonon dispersion is calculated by using an interatomic pair potential model within General Utility Lattice Program (GULP)[25].

From the measured thermal conductivity, one can reliably estimate the average mean-free-path from the measured conductivity. From Eq. (7b), we can write

$$\langle\langle \lambda_{ph} \rangle\rangle = \frac{\kappa_{ph}}{\frac{k_B^2 T_L \pi^2}{3h} \int_{-\infty}^{+\infty} \frac{M_{ph}(\omega)}{A}\left(\frac{3}{\pi^2}\left(\frac{\hbar\omega}{k_B T_L}\right)^2\left(-\frac{\partial n_0}{\partial(\hbar\omega)}\right)\right)d(\hbar\omega)} = \frac{\kappa_{ph}}{K_{ph\_BAL}/A}. \quad (9)$$

where the denominator of Eq. (9) is recognized as ballistic thermal conductance per area $K_{ph\_BAL}/A$. Note the units of $K_{ph\_BAL}/A$ are $W/m^2$-$K$, which is different from the units of $\kappa_{ph}$, $W/m$-$K$. Since the numerator in Eq. (9) can be measured and the denominator readily evaluated from a known dispersion, reliable estimates of the mean-free-paths can be obtained. The extraction of average electron mean-free-path by similar way has been used to analyze electronic devices[26-28]

Finally, we note that two Debye temperatures are required to evaluate the specific heat and the thermal conductivity properly with the simple Debye model. The use of two Debye temperatures was also found necessary in work on the thermal conductivity of nanowires[18]. The Debye temperature is usually determined to obtain the observed specific heat, which is hereinafter called Debye temperature for the specific heat ($\Theta_D$). The Debye temperature for thermal conductivity ($\Theta_M$) is newly defined and is determined to obtain the correct $K_{ph\_BAL}$. For 3D bulk ballistic conductors with linear phonon dispersion, the specific heat per volume and ballistic thermal conductance per area are expressed as

$$C_{V,3D} = \left(\frac{3k_B^4 T_L^3}{2\pi^2 \hbar^3 \upsilon_s^3}\right) \int_0^{\Theta_D/T} \frac{x^4 e^x}{(e^x-1)^2} dx \qquad (10a)$$

$$K_{ph\_BAL}/A = \left(\frac{3k_B^4 T_L^3}{8\pi^2 \hbar^3 \upsilon_s^2}\right) \int_0^{\Theta_M/T} \frac{x^4 e^x}{(e^x-1)^2} dx \qquad (10b)$$

where $x = \hbar\omega/k_B T_L$. Both $\Theta_D$ and $\Theta_M$ are extracted to match full band results.

### 3) Results

In this section, the phonon thermal conductivity will be evaluated and interpreted within the Landauer framework. Two representative semiconductor materials are compared to examine key factors for good TE materials; Si and $Bi_2Te_3$. We then compare full band calculations to linear dispersion approximations. A technique to extract a well-defined average mean-free-path is also presented. We show that this mean-free-path has a strong physical significance.

Figures 1(a) and 1(b) show calculated and measured[29-31] phonon dispersion characteristic along high symmetry directions; the phonon DOS for Si and $Bi_2Te_3$ are also shown. The Tersoff potential model and Morse potential model are used for Si[32] and $Bi_2Te_3$[24], respectively. Computed elastic properties with these models show overall good agreement with experiments, indicating that the potential model describe well the harmonic behavior[24,32]. For both materials, it can be seen that the AP modes are well reproduced and OP modes are somewhat overestimated. Because thermal transport is mostly dominated by AP modes, we expect that these full band dispersions will predict the lattice thermal conductivity well. It should be understood that our objective is to describe a general technique and to discuss general features of the solution. More refined treatments of phonon dispersion could be used.

The specific heat per volume (the integral of Eq. (A3)) is calculated and shown in Fig. 1(c). As shown in Eq. (A3), the specific heat calculations do not require us to include scattering. In each plot, the solid line is the result with calculated DOS, and the dashed line is the result with the measured DOS. It is seen that two curves are generally in a good agreement for both materials.

Figure 2(b) shows full dispersion results and results with a linear dispersion approximation. At low frequency, the linear dispersion approximation provides a good fit to the full band calculation, and it is found that the average sound velocity to fit to full band results for $Bi_2Te_3$ ( $v_s = 1.74 \times 10^5 \, cm/s$ ) is about one third of average sound velocity of Si ( $v_s = 5.32 \times 10^5 \, cm/s$ ). The available number of conducting modes is seen to be smaller for $Bi_2Te_3$ for most of frequency range.

Next, the phonon thermal conductance is evaluated. The ballistic thermal conductance per area for Si and $Bi_2Te_3$ is calculated as shown in Fig. 3. The ballistic thermal conductance is proportional to the effective number of phonon conducting modes, which can be readily obtained from phonon dispersions. Below 30 K, $K_{ph\_BAL}$ for $Bi_2Te_3$ is larger than Si due to the large $M_{ph}(\omega)$ at low frequency. At 300 K, however, $K_{ph\_BAL}$ for Si is a factor of 10 larger than $K_{ph\_BAL}$ for $Bi_2Te_3$, which results because the effective number of phonon conducting modes of $Bi_2Te_3$ is 10 times smaller than that of Si.

From Eq. (9), the average mean-free-path for backscattering is deduced by taking ratio of the measured thermal conductivity to the ballistic thermal conductance per area. Figure 4 shows the extracted $\langle\langle\lambda_{ph}\rangle\rangle$ for Si and $Bi_2Te_3$. At 300 K, $\langle\langle\lambda_{ph}\rangle\rangle$ for $Bi_2Te_3$ is 14 nm and for Si, it is 115 nm. To relate extracted $\langle\langle\lambda_{ph}\rangle\rangle$ to expected average $\langle\langle\lambda_{ph}\rangle\rangle$ from the spectral phonon mean-free-path for backscattering $\lambda_{ph}(\omega)$, expressions for $\tau_{ph}(\omega)$ in the relaxation time approximation (RTA) are used for umklapp[33], point defect[34], and boundary[35] scattering rate: $\tau_u^{-1} = B\omega^3 e^{-C/T}, \tau_d^{-1} = D\omega^4,$ and $\tau_b^{-1} = \langle\upsilon(\omega)\rangle/(F \cdot l)$, respectively. The parameters to fit extracted $\langle\langle\lambda_{ph}\rangle\rangle$ are $B = 2.8 \times 10^{-19}$ s/K, $C = 140$ K, $D = 1.32 \times 10^{-45}$ $s^3$, $F = 0.4$, $l = 7.16 \times 10^{-3}$ m for Si and $B = 2.8 \times 10^{-18}$ s/K, $C = 10$ K, $D = 1.32 \times 10^{-45}$ $s^3$, $F \cdot l = 1 \times 10^{-4}$ m for $Bi_2Te_3$. As shown in the inset of Fig. 3, experimental thermal conductivity[35,36] and calculations are in a good agreement, which indicate that the extracted $\langle\langle\lambda_{ph}\rangle\rangle$ has strong physical significance.

In the Landauer picture, the low thermal conductivity of $Bi_2Te_3$ at 300 K is attributed to two factors. First of all, it has an effective number of conduction channels that is ten times smaller than Si, as shown in Fig. 3. The different number of conducting channels are related to the different phonon dispersions. On top of that, $Bi_2Te_3$ has a smaller $\langle\langle\lambda_{ph}\rangle\rangle$ due to umklapp scattering which is a factor of 10 stronger than for Si. Both factors lead to two orders of magnitude reduction in thermal conductivity comparing to Si. For comparison, average electron mean-free-paths at room temperature are 18 nm for $Bi_2Te_3$ and 13 nm for Si, which are extracted in a similar way from the full band electronic structure[19]. The number of electron conducting modes for Si and $Bi_2Te_3$ are shown in Fig. 7(b). In terms of electronic performance, the effective number of conduction channels and average electron mean-free-path for backscattering are similar for both materials, resulting in similar power factor($S^2G$) value.

**4) Discussion**

In this section, we show that average mean-free-path obtained by simple estimates differs by an order of magnitude from that extracted from full phonon dispersion. The use of simplified dispersion model is then discussed.

The simplest conventional approach to estimate bulk mean-free path from a classical kinetic theory is

$$\langle\langle \Lambda \rangle\rangle = 3\kappa_{ph}/C_V \upsilon_s \quad , \tag{11}$$

where $\kappa_{ph}$, $\upsilon_s$, and $C_V$ are measured quantities. Using $\upsilon_s = 2.95 \times 10^5$ cm/s and $C_V = 1.20 \times 10^6$ J/cm$^3$-K from Fig. 1(b) and (c) for Bi$_2$Te$_3$, estimated $\langle\langle \Lambda \rangle\rangle$ at 300 K is 1.2 nm, an order of magnitude smaller than 14 nm extracted from Eq. (9). This occurs because the appropriate velocity we should use in Eq. (11) is much different from the measured sound velocity ($\upsilon_s = 2.95 \times 10^5$ cm/s). Equation (A3) can be re-arranged as

$$\kappa_{ph} = \frac{1}{3} C_V \upsilon_{ave} \langle\langle \Lambda \rangle\rangle \tag{12}$$

where $\upsilon_{ave} \left( = \int d(\hbar\omega) C_V(\omega) \upsilon(\omega) / C_V \right)$ is the appropriate average velocity we should use in Eq. (12) and $\langle\langle \Lambda \rangle\rangle \left( = (3/4) \langle\langle \lambda \rangle\rangle \right)$ is an average mean-free-path for scattering which is different from $\langle\langle \lambda_{ph} \rangle\rangle$ as is seen in Eq. (6). At 300 K, $\upsilon_{ave} = 3.40 \times 10^4$ cm/s, a 5× ~ 9× smaller than either the measured sound velocity ($\upsilon_s = 2.95 \times 10^5$ cm/s) or the average sound velocity to fit to full band results ($v_s = 1.74 \times 10^5$ cm/s). Therefore, prediction of the bulk mean-free path from measured sound velocity in Eq. (11) leads to serious errors.

Simple phonon dispersion models are often used to analyze thermoelectric devices. The simplest model is the Debye model (the linear dispersion approximation) with Debye temperature ($\Theta_D$). As mentioned in Sec. 2, a Debye temperature for thermal conductivity ($\Theta_M$) must be newly defined to treat the thermal conductivity properly with the simple Debye model. Both $\Theta_D$ and $\Theta_M$ are extracted to match full band results as shown in Fig. 5 (a) for Si and Bi$_2$Te$_3$. It can be seen that $\Theta_M$ is smaller than $\Theta_D$ by 20~50 % depending on materials and the two Debye temperatures are weakly dependent on temperature. For Bi$_2$Te$_3$, Fig. 5(b) compare the full band and Debye models of $M_{ph}(\omega)$ and $D_{ph}(\omega)$. The Debye cutoff frequencies, $\omega_D (= k_B \Theta_{D,\max}/\hbar)$ and $\omega_M (= k_B \Theta_{M,\max}/\hbar)$ are 3.57 *THz* and 1.93 *THz*, respectively.

Next, the specific heat and thermal conductivity are calculated as shown in Fig. 6(a) and the inset of Fig. 6(a). Although only the Debye cutoff frequency from the phonon density of states ($\omega_D$) gives the correct specific heat, thermal conductivities obtained from either of the two Debye frequencies, $\omega_M$ and $\omega_D$, match well the results using full phonon dispersion. Use of the Debye model with the cutoff frequency, $\omega_D$, however, leads to serious errors in estimating $\langle\langle \lambda_{ph} \rangle\rangle$ — by one order of magnitude as shown in Fig. 6(b). The average mean-free-path obtained from Debye model with the cutoff frequency, $\omega_M$ is in relatively a good agreement with that obtained by full phonon dispersion because this cutoff frequency gives the correct effective number of conduction channels, i.e. the ballistic thermal conductance per area. The inset of Fig. 6(b), however, shows that the spectral mean-free-path, $\lambda_{ph}(\omega)$ obtained from Debye models doesn't capture exactly the detailed frequency dependence of $\lambda_{ph}(\omega)$ from full phonon dispersion – regardless of choice of cutoff frequency.

The reason that an effective mass description works well for electrons and a Debye model does not works as well for phonon is that the important energy for electrons is near the bottom of the band, but for phonons it is the entire phonon dispersion. Fig. 7(a) shows the "window" function for phonons, $W_{ph}$, and number of phonon conducting modes for Si and $Bi_2Te_3$. Comparing to the "window" function for electrons, $W_{el}$, and number of electron conducting modes as shown in Fig. 7(b), it can be easily seen that the entire frequency range of full phonon dispersion affect thermal conductivity for Si and $Bi_2Te_3$. However, only the electron dispersion around the band edges is important. Note that this occurs because of difference between full phonon spectrum and full electronic structure, not because of difference between Fermi-Dirac and Bose-Einstein distribution or between "window" functions, defined by Eqs. (7e) and (7f). As shown in Fig. 7, both "window" functions, $W_{ph}$ and $W_{el}$, have a width of ~ 5 $k_B T_L$ and the function $W_{ph}$ has similar energy dependence to the so-called Fermi "window" function $W_{el}$. The integral of window functions for electrons, $W_{el}$ from -$\infty$ to $\infty$ gives 1, while the integral of $W_{ph}$ from 0 to $\infty$ is 1. This comparison explains why effective mass approximation (EMA) works well for electron transport[19], and Debye model should be used with caution.

5) **Summary and Conclusion**

In this paper, we related the Landauer approach for phonon transport to the more commonly used Boltzmann transport equation approach. Although the Landauer approach applies from the ballistic to diffusive limit and for 1D, 2D, and 3D conductors, we restricted our attention in this paper to the diffusive limit and to 3D, bulk materials. The common expression for thermal conductance that begins with classical kinetic theory was related to the corresponding Landauer expression. A simple "counting bands" technique for extracting the kernel of the transport integral, $M(E)$, was illustrated. As an example of the technique, we examined the electronic and thermal performance of Si and $Bi_2Te_3$ using a full band description of phonon dispersion and electronic bandstructure. A simple technique for extracting a physically well defined mean-free-path for phonons was presented. This mean-free-path agrees with a simple estimate from the measured specific heat – as long as the appropriate average velocity obtained from the given phonon dispersion is used. Finally, we discussed the use of simple phonon dispersion models, such as the widely used Debye model for phonon dispersion, which is widely used for device design. We showed that two different Debye temperatures are needed – one to describe the phonon density of states and specific heat and another to describe the distribution of conducting channels, $M(E)$, and the thermal conductivity. The existence of two different Debye temperatures is analogous to the two effective masses needed to describe electron transport, the conductivity and density-of-states effective masses. Using the conductivity Debye temperature and the measured lattice thermal conductivity, a physically meaningful average mean-free-path can be accurately obtained. Finally, we explained why the effective mass model works well for electrons and why the Debye model does not work as well for phonons. Although the conclusion is that simplified phonon models should be used with caution, the simple procedure for evaluating $M(E)$ from the full phonon dispersion provides a practical alternative.

**Acknowledgments**

This work was supported by MARCO Materials Structures and Devices (MSD) Focus Center as well as the Network for Computational Nanotechnology (NCN) for computational resources. C. J. thanks Bo Qiu and Dhrub Singh for insightful discussions.

**Appendix**

To relate Eq. (5b), the Landauer expression for lattice thermal conductivity, to the conventional expression from kinetic theory, we write the number of phonon conduction channels per area, $M'_{ph}(\omega)$, as [19]

$$M'_{ph}(\omega) = (h/4)\upsilon(\omega)D'_{ph}(\omega)  \tag{A1}$$

where $D'_{ph}(\omega) \equiv (1/\Omega)\sum_q \delta(\hbar\omega - \hbar\omega_q)$ is the phonon density of states (DOS) per polarization per volume. Using Eqs. (6) and (7), Eq. (5b) the phonon thermal conductivity can be written in the conventional form[37,38] as

$$\kappa_{ph} = \frac{1}{3}\int_{-\infty}^{+\infty} d(\hbar\omega) C_V(\omega)\upsilon_{ph}(\omega)\Lambda_{ph}(\omega) \tag{A2}$$

with $C_V(\omega)$ being the specific heat per unit volume

$$C_V(\omega) = k_B^2 T_L D_{ph}(\omega)\left(\frac{\hbar\omega}{k_B T_L}\right)^2 \left(-\frac{\partial n_0}{\partial(\hbar\omega)}\right). \tag{A3}$$

Next, for both electron and phonon, the number of conduction channels is defined as[19]

$$M(E) = \frac{h}{2L}\sum_k |\upsilon_x|\delta(E - E_k) \tag{A4}$$

Assuming parabolic dispersion for electron ($E - \varepsilon_1 = \hbar^2 k^2/2m_e$), corresponding expression for the number of conduction channels per spin per valley for 1D, 2D, and 3D conductors are given as[20]

$$M_{el}(E) = \Theta(E - \varepsilon_1) \quad \text{(1D)} \tag{A5a}$$

$$M_{el}(E) = W\frac{\sqrt{2m_e(E - \varepsilon_1)}}{\pi\hbar} \quad \text{(2D)} \tag{A5b}$$

$$M_{el}(E) = A\frac{m_e}{2\pi\hbar^2}(E - E_C) \quad \text{(3D)} \tag{A5c}$$

where $\Theta$ is the unit step function, $\varepsilon_1$ is the bottom of the first subband, $m_e$ is the electron effective mass, $E_C$ is the conduction band edge and $W$ and $A$ are the width and the area of the 2D and 3D conductors, respectively. For phonons with linear and isotropic dispersion approximation, $\omega = \upsilon_s q$, $M_{ph}(\omega)$ per polarization is given as

$$M_{ph}(\omega) = \Theta(\omega) \quad \quad (1D) \quad \quad \text{(A6a)}$$

$$M_{ph}(\omega) = W(\omega/\pi\upsilon_s) \quad \quad (2D) \quad \quad \text{(A6b)}$$

$$M_{ph}(\omega) = A(\omega^2/4\pi\upsilon_s^2) \quad \quad (3D) \quad \quad \text{(A6c)}$$

where $\upsilon_s$ is the velocity of sound.

**References:**

[1] A. Majumdar, Science **303**, 777-778 (2004).
[2] G. J. Snyder and E. S. Toberer, Nat. Mater. **7**, 105-114 (2008).
[3] G. Chen, M. S. Dresselhaus, G. Dresselhaus, J. P. Fleurial, and T. Caillat, Int. Mater. Rev. **48**, 45 (2003).
[4] J. P. Heremans, V. Jovovic, E. S. Toberer, A. Saramat, K. Kurosaki, A. Charoenphakdee, S. Yamanaka, and G. J. Snyder, Science **321**, 554-557 (2008).
[5] T. Koga, X. Sun, S. B. Cronin, and M. S. Dresselhaus, Appl. Phys. Lett. **73**, 2950-2952 (1998).
[6] T. Thonhauser, T. J. Scheidemantel, J. O. Sofo, J. V. Badding, and G. D. Mahan, Phys. Rev. B **68**, 85201 (2003).
[7] C. M. Finch, V. M. Garcia-Suarez, and C. J. Lambert, Phys. Rev. B **79**, 033405-4 (2009).
[8] D. Dragoman and M. Dragoman, Appl. Phys. Lett. **91**, 203116-3 (2007).
[9] A. Bentien, S. Johnsen, G. K. H. Madsen, B. B. Iversen, and F. Steglich, Europhys. Lett. **80**, 17008 (2007).
[10] E. Pop, S. Sinha, and K. E. Goodson, Proceedings of the IEEE **94**, 1587-1601 (2006).
[11] M. Lundstrom, *Fundamentals of Carrier Transport,* (Cambridge University Press, 2000).
[12] G. Chen, *Nanoscale Energy Transport and Conversion,* (Oxford University Press, 2005).
[13] Supriyo Datta, *Electronic Transport in Mesoscopic Systems,* (Cambridge University Press, 1997).
[14] D. E. Angelescu, M. C. Cross, and M. L. Roukes, Superlattice. Microst. **23**, 673-689 (1998).
[15] B. J. Van Wees, H. Van Houten, C. W. J. Beenakker, J. G. Williamson, L. P. Kouwenhoven, D. Van der Marel, and C. T. Foxon, Phys. Rev. Lett. **60**, 848-850 (1988).
[16] D. A. Wharam, T. J. Thornton, R. Newbury, M. Pepper, H. Ahmed, J. E. F. Frost, D. G. Hasko, D. C. Peacock, D. A. Ritchie, and G. A. C. Jones, J. of Phys. C: Solid State Phys. **21**, L209 (1988).
[17] Luis G. C. Rego and G. Kirczenow, Phys. Rev. Lett. **81**, 232 (1998).
[18] N. Mingo, Phys. Rev. B **68**, 113308 (2003).
[19] C. Jeong, R. Kim, M. Luisier, S. Datta, and M. Lundstrom, J. of Appl. Phys. **107**, 023707 (2010).
[20] R. Kim, S. Datta, and M. S. Lundstrom, J. Appl. Phys. **105**, 034506-6 (2009).
[21] G. D. Mahan and J. O. Sofo, Proc. Natl. Acad. Sci. **93**, 7436-7439 (1996).
[22] N. Mingo, Phys. Rev. B **68**, 113308 (2003).
[23] B.-L. Huang and M. Kaviany, Phys. Rev. B **77**, 125209 (2008).
[24] B. Qiu and X. Ruan, Phys. Rev. B **80**, 165203 (2009).
[25] J. D. Gale and A. L. Rohl, Mol. Simulat. **29**, 291-341 (2003).
[26] J.-Y. Park, S. Rosenblatt, Y. Yaish, V. Sazonova, H. Üstünel, S. Braig, T. A. Arias, P. W. Brouwer, and P. L. McEuen, Nano Letters **4**, 517-520 (2004).
[27] K. I. Bolotin, K. J. Sikes, J. Hone, H. L. Stormer, and P. Kim, Phys. Rev. Lett. **101**, 096802-4 (2008).
[28] C. Jeong, D.A. Antoniadis, and M.S. Lundstrom, IEEE Trans. Electron Devices, **56**, 2762-2769 (2009).
[29] P. A. Temple and C. E. Hathaway, Phys. Rev. B **7**, 3685 (1973).
[30] H. Rauh, R. Geick, H. Kohler, N. Nucker, and N. Lehner, J. of Phys. C: Solid State Phys. **14**, 2705 (1981).
[31] J. O. Jenkins, J. A. Rayne, and R. W. Ure, Phys. Rev. B **5**, 3171 (1972).
[32] J. Tersoff, Phys. Rev. B **39**, 5566 (1989).
[33] M. Asen-Palmer, K. Bartkowski, E. Gmelin, M. Cardona, A. P. Zhernov, A. V. Inyushkin, A. Taldenkov, V. I. Ozhogin, K. M. Itoh, and E. E. Haller, Phys. Rev. B **56**, 9431 (1997).
[34] P. G. Klemens, Proc. Phys. Soc. **A68**, 1113 (1955).
[35] M. G. Holland, Phys. Rev. **132**, 2461 (1963).
[36] P. A. Walker, Proc. Phys. Soc. **76**, 113-126 (1960).
[37] J. Callaway, Phys. Rev. **113**, 1046 (1959).
[38] N. W. Ashcroft and N. D. Mermin, *Solid State Physics*, 1st ed. (Brooks Cole, 1976).

**Figure Captions**

Fig. 1. Phonon dispersion along the high symmetry lines and phonon density of states for (a) Si and (b) $Bi_2Te_3$. (c) Specific heat per volume of Si (Inset) and $Bi_2Te_3$. The solid and dashed lines are for the results of calculation and experiments.

Fig. 2. (a) Illustration of bands counting method for specific dispersion relation for Si. Dotted line is guide to eye. (b) Number of phonon modes of Si and $Bi_2Te_3$. The dashed and dashed-dot lines are the results obtained from the Debye model with the fitted sound velocities of $1.74 \times 10^5$ cm/s and $5.32 \times 10^5$ cm/s for Si and $Bi_2Te_3$, respectively.

Fig. 3. Ballistic thermal conductance per area of Si and $Bi_2Te_3$. Insets: lattice thermal conductivity of Si and $Bi_2Te_3$. The solid line and symbols are calculated and experimental values, respectively.

Fig. 4. Extracted average mean-free-path of Si and $Bi_2Te_3$ by taking the ratio of experimental thermal conductivity to ballistic thermal conductance per area.

Fig. 5. (a) Comparison of Debye temperature for the specific heat ($\Theta_D$) and Debye temperature for thermal conductivity ($\Theta_M$) normalized by maximum $\Theta_D$. (b) Debye model vs. full dispersion for $Bi_2Te_3$. Red and green solid lines are full band results of number of phonon modes ($M_{ph}$) and density of states ($D_{ph}$) in arbitrary units. The dashed-dot and dashed lines are results from Debye approximation at low frequency for $M_{ph}$ and $D_{ph}$.

Fig. 6. (a) specific heat calculation and (b) extracted average mean-free-path of $Bi_2Te_3$ for Debye models and full band results. Insets of Fig. 6(a) and Fig. 6(b) are thermal conductivity calculation and spectral mean-free-path for backscattering, respectively. Solid lines: full band results; Dashed and dashed dot lines: Debye approximation.

Fig. 7. (a) number of phonon conducting modes ($M_{ph}$) and Eq. (7f) at 300 K, (b) number of electron conducting modes ($M_{el}$) calculated from full band electronic structure and Eq. (7e) at 300 K. For horizontal axis, $\varepsilon = E_{el} - E_C$ for number of electron conducting modes and $\varepsilon = E_{el} - E_F$ for a electron "window function" assuming $E_F = E_C$ which is a typical condition for optimum performance.

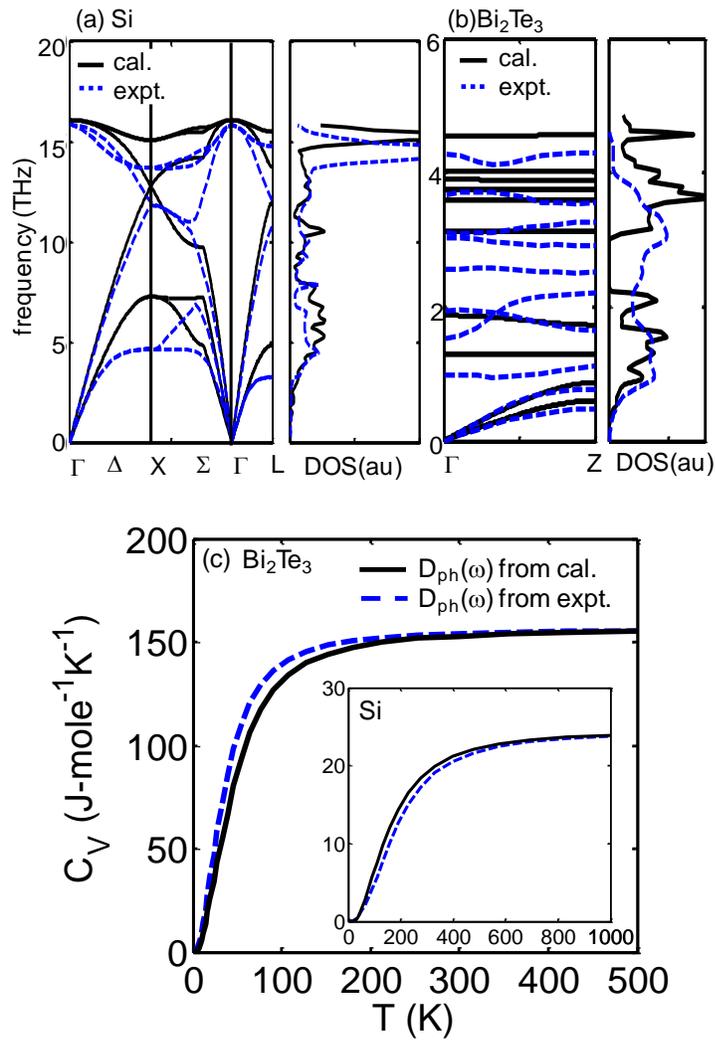

Fig. 1. Phonon dispersion along the high symmetry lines and phonon density of states for (a) Si and (b) $Bi_2Te_3$. (c) Specific heat per volume of Si (Inset) and $Bi_2Te_3$. The solid and dashed lines are for the results of calculation and experiments.

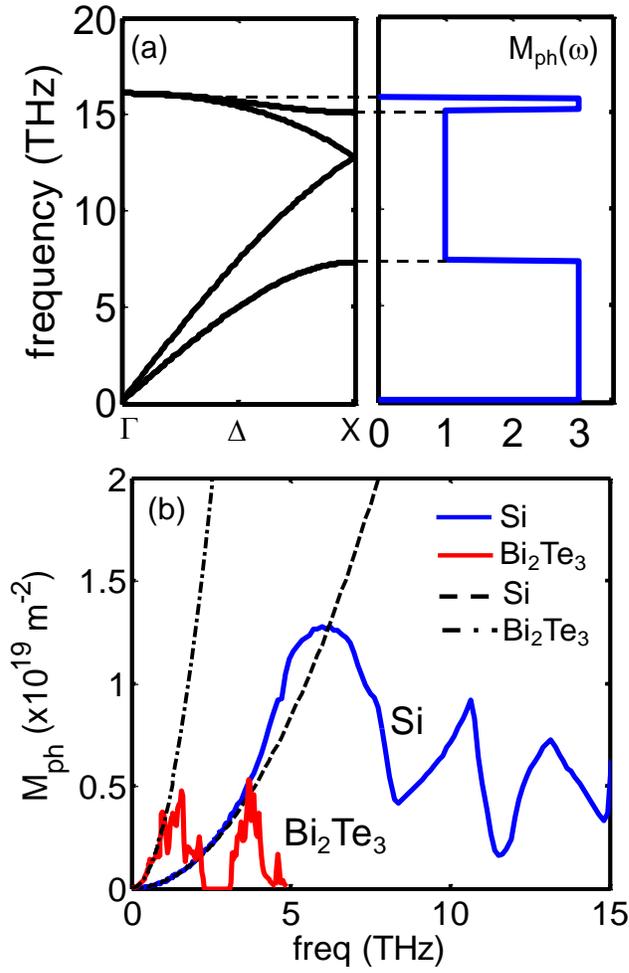

Fig. 2. (a) Illustration of bands counting method for specific dispersion relation for Si. Dotted line is guide to eye. (b) Number of phonon modes of Si and $Bi_2Te_3$. The dashed and dashed-dot lines are the results obtained from the Debye model with the fitted sound velocities of $1.74 \times 10^5$ cm/s and $5.32 \times 10^5$ cm/s for Si and $Bi_2Te_3$, respectively.

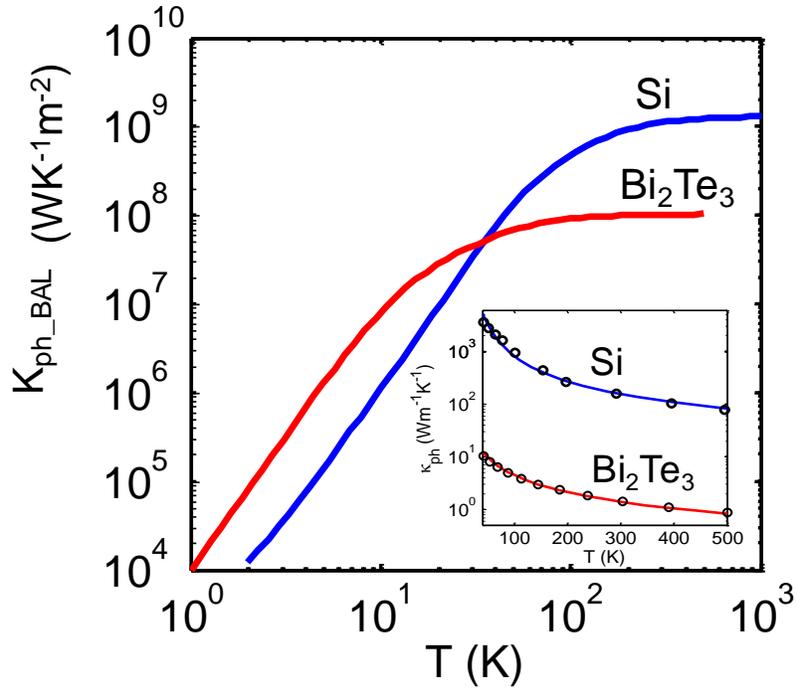

Fig. 3. Ballistic thermal conductance per area of Si and $Bi_2Te_3$. Insets: lattice thermal conductivity of Si and $Bi_2Te_3$. The solid line and symbols are calculated and experimental values, respectively.

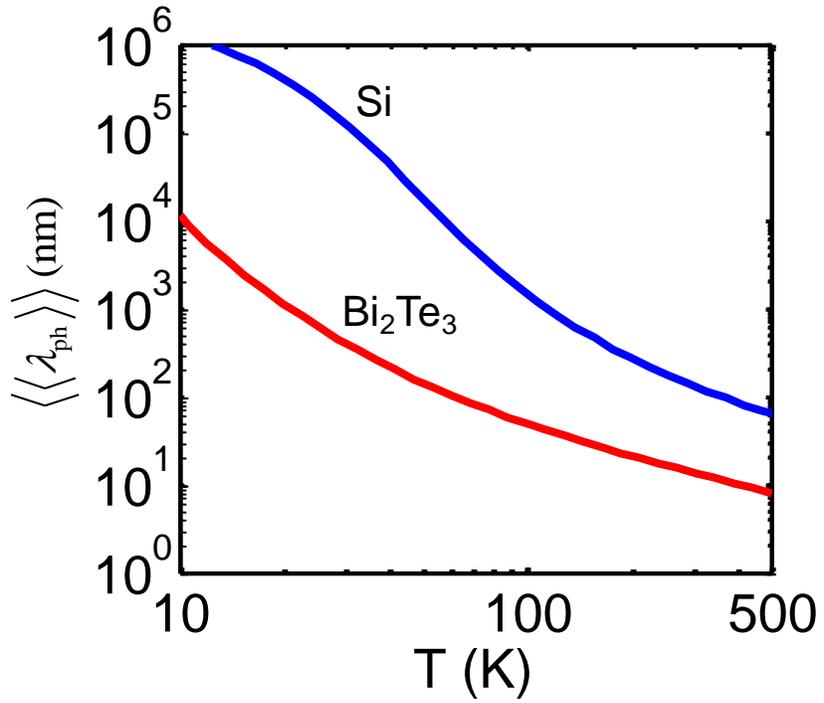

Fig. 4. Extracted average mean-free-path of Si and $Bi_2Te_3$ by taking the ratio of experimental thermal conductivity to ballistic thermal conductance per area.

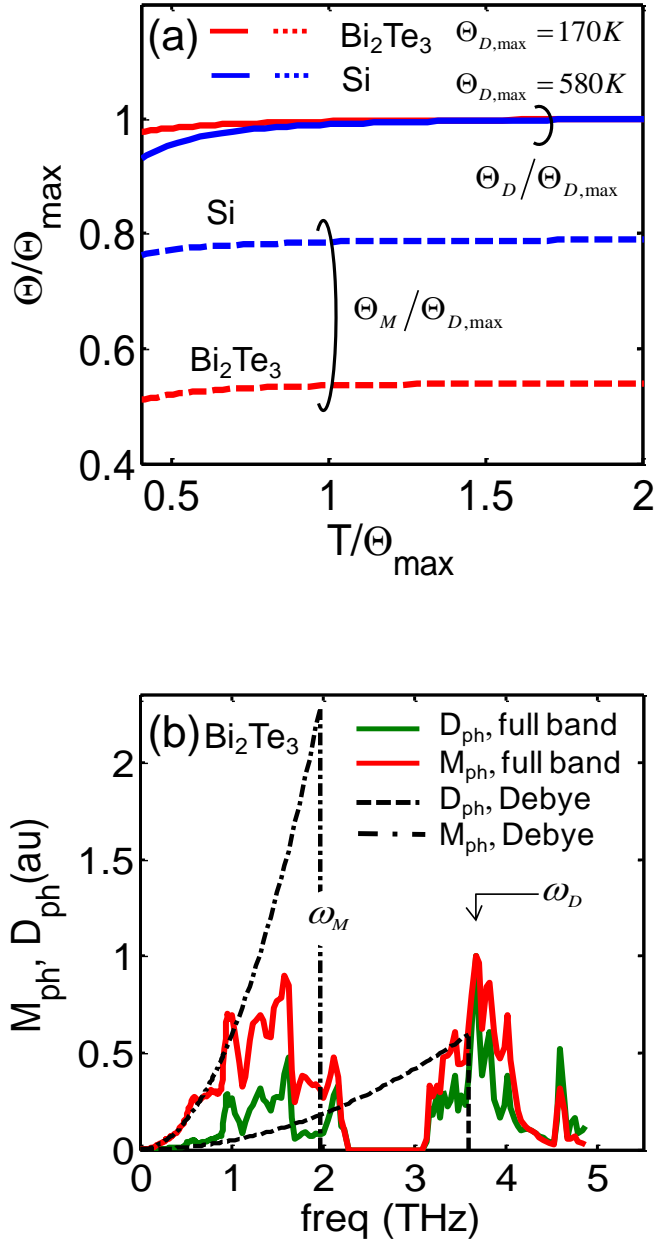

Fig. 5. (a) Comparison of Debye temperature for the specific heat ($\Theta_D$) and Debye temperature for thermal conductivity ($\Theta_M$) normalized by maximum $\Theta_D$. (b) Debye model vs. full dispersion for $Bi_2Te_3$. Red and green solid lines are full band results of number of phonon modes ($M_{ph}$) and density of states ($D_{ph}$) in arbitrary units. The dashed-dot and dashed lines are results from Debye approximation at low frequency for $M_{ph}$ and $D_{ph}$.

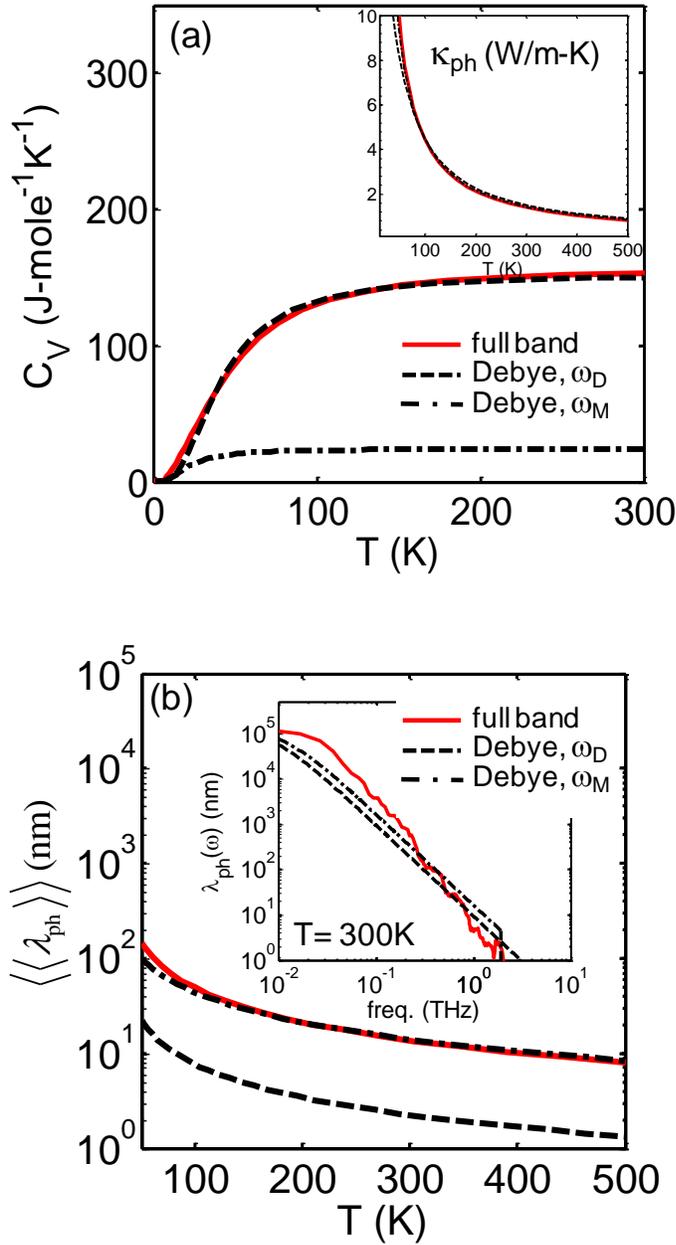

Fig. 6. (a) specific heat calculation and (b) extracted average mean-free-path of $Bi_2Te_3$ for Debye models and full band results. Insets of Fig. 6(a) and Fig. 6(b) are thermal conductivity calculation and spectral mean-free-path for backscattering, respectively. Solid lines: full band results; Dashed and dashed dot lines: Debye approximation.

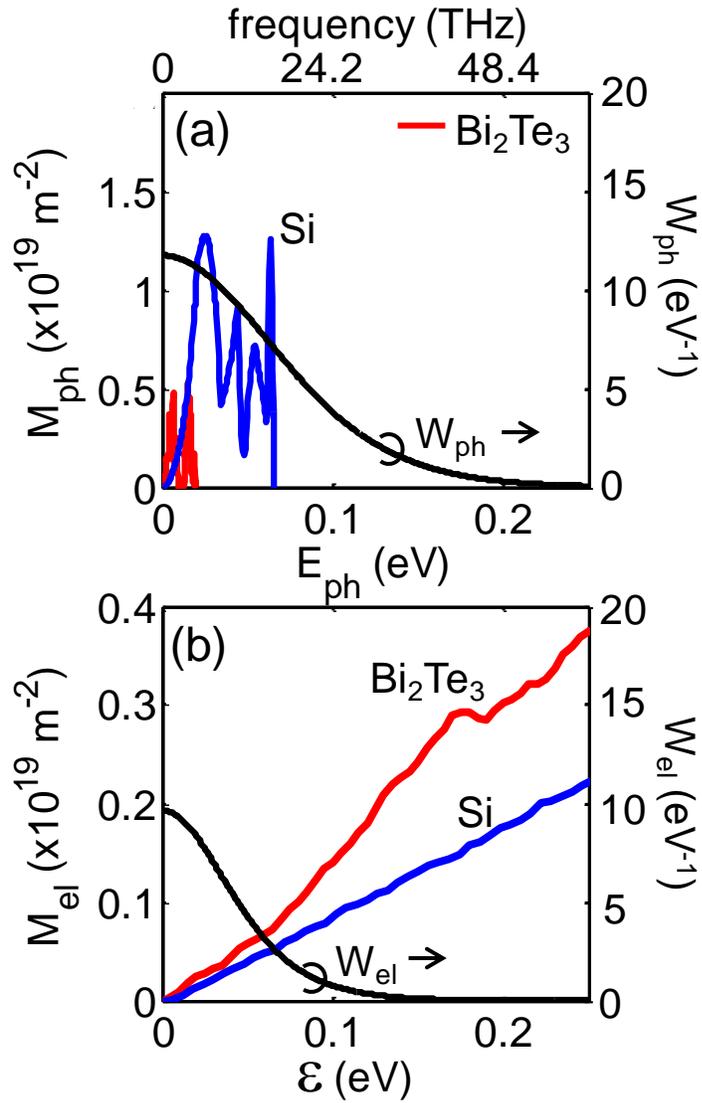

Fig. 7. (a) number of phonon conducting modes ($M_{ph}$) and Eq. (7f) at 300 K, (b) number of electron conducting modes ($M_{el}$) calculated from full band electronic structure and Eq. (7e) at 300 K. For horizontal axis, $\varepsilon = E_{el} - E_C$ for number of electron conducting modes and $\varepsilon = E_{el} - E_F$ for a electron "window function" assuming $E_F = E_C$ which is a typical condition for optimum performance.